\documentclass[aps,12pt,superscriptaddress,tightenlines,preprint,floatfix]{revtex4-1}
\usepackage[dvips]{graphicx}
\usepackage{amsmath,amssymb,amsthm,bm}

\newcommand{\R}{\mathbb{R}}
\DeclareMathOperator{\id}{id}
\DeclareMathOperator{\im}{im}

\DeclareMathOperator{\linearspan}{span}

\allowdisplaybreaks

\begin{document}
\title{New invariants for entangled states}
\author{Roman V. Buniy}
\email{roman.buniy@gmail.com}
\affiliation{School of Earth and Space Exploration, Arizona State University, Tempe, AZ 85287}
\affiliation{Chapman University, Schmid College of Science, Orange, CA 92866}
\altaffiliation{Permanent address}
\author{Thomas W. Kephart}
\email{tom.kephart@gmail.com}
\affiliation{Department of Physics and Astronomy, Vanderbilt University, Nashville, TN 37235}
\date{\today}
\begin{abstract}
  We propose new algebraic invariants that distinguish and classify entangled states.
  Considering qubits as well as higher spin systems, we obtained complete entanglement classifications for cases that were either unsolved or only conjectured in the literature.
\end{abstract}
\maketitle

\section{Introduction}

Ever since the formulation of the EPR paradox \cite{EPR}, the phenomenon of quantum entanglement has generated an ever increasing stream of research.
Although applications of entangled states are already advanced and extensive, fundamental understanding of the phenomenon is nevertheless still incomplete.
Such an understanding requires the description of qualitative features of entanglement (which are well-known and simple to formulate) and its quantitative features (which are very complex and not completely understood).
The latter are usually described in terms of entanglement invariants that can distinguish entangled states, and finding these is very difficult.
Since distinction implies classification, this leads to the problem of entanglement classification, which is a prominent unsolved problem in quantum information theory.

Entanglement has been studied mostly for subsystems of spin one half.
Quantum computation favors this case since a qubit with its two degrees of freedom represents the fundamental unit of quantum information.
However, as we do not yet know on which quantum systems large practical quantum computers will ultimately be based, higher spin quantum states should be studied as well.
Such a study would be much advanced by understanding the classification of entangled states for the general case of subsystems of arbitrary spins.

The classical theory of invariants \cite{Olver} provides the standard method of finding entanglement invariants.
Variants of this method are used in most known cases of partial or complete classification; see, for example, \cite{Gelfand,Dur,Verstraete,Klyachko,Miyake1,Luque1,Miyake2,Briand,Luque2,Toumazet,Cao,Lamata,Borsten:2008wd,Borsten:2010db}.
We propose a new method of entanglement classification that uses only basic linear algebra \cite{algebra} and whose invariants are discrete algebraic invariants complementing the known continuous invariants.

The following brief review of key properties of entangled states motivates the use of algebraic invariants for entanglement classification.
In the main text we fully develop these ideas and apply them to several cases of three entangled higher spin systems, only some of which are known in the literature.

We first observe that the phenomenon of entanglement is a consequence of the superposition principle and the tensor product postulate.
By the superposition principle, the state space of a physical system is a vector space, so that any linear combination of state vectors is such as well.
By the tensor product postulate, the state space of a system consisting of several subsystems is a subspace of the tensor product of the state spaces of all the subsystems.
The principle and postulate together imply that a state vector of the system is a linear combination of tensor products of state vectors of the subsystems.

Consequently, the information provided by a given state of the system can be divided into two parts: (1) a list of contributing states of the subsystems and (2) the manner in which these are combined.
The former and the latter describe respectively macro and micro properties of the states, and entanglement characterizes the latter.
For given states of the subsystems (the first part is fixed), there are various ways in which linear combinations of their tensor products can be formed (the second part varies).
In the simplest case only one state from each subsystem contributes, and there is only one term in the linear combination, which results in no entanglement.
As numbers of contributing states from each subsystem increase, new ways to form linear combinations become available.
If new added states are linearly independent from those already included, this results in states that have, in general, higher degrees of entanglement.
The process rapidly becomes complicated for a large number of subsystems because of the combinatorial nature of the procedure.
Exploring all resulting possibilities and partitioning states into corresponding classes formed by related states is the goal of entanglement classification.

We now proceed with explicit details of our method and demonstrate its use with several examples.
A more general, comprehensive, and detailed discussion of the method---including a theorem on a correspondence between entanglement classes and algebraic invariants---and its application to more complicated examples can be found in \cite{Buniy:2010zp}.

\section{Method}

Let $S$ be a system that consists of subsystems $S_1,\dotsc,S_n$, and assume that state spaces of $S,S_1,\dotsc,S_n$ are finite dimensional vector spaces $V,V_1,\dotsc,V_n$, respectively.
(We consider vector spaces over $\R$ for the simplicity of presentation; the case of complex vector spaces needs only minimal modifications.)
The space $V$ is a subspace of $V_1\otimes\dotsb\otimes V_n$, where a specific choice of $V$ is determined  by the nature of the subsystems.
In a particularly important case of identical subsystems, $V$ is determined by a permutation symmetry acting on the subsystems.
We consider here only the simplest case where $V=V_1\otimes\dotsb\otimes V_n$.

Entanglement properties of $v\in V$ are determined by specific ways in which $v$ is formed from elements of $V_1,\dotsc,V_n$.
From this point of view, the simplest elements of $V$ are decomposable vectors.
Any decomposable vector $v\in V$ can be written in the factorizable form $v=v_1\otimes\dotsb\otimes v_n$, where $v_i\in V_i$, and an important property of such a vector $v$ is that each subsystem $S_i$ is in a definite state $v_i$.
Although decomposable vectors comprise only a small part of $V$, they span all of it.
This simple property of tensor products leads to remarkable complications and plays the central role in our classification of entangled states.

Nondecomposable vectors are vectors that cannot be written in the factorizable form, and for such states, we cannot say in which state each subsystem is.
The simplest example of a nondecomposable vector in $V$ is $v+v'$, where $v=v_1\otimes\dotsb\otimes v_n$, $v'=v'_1\otimes\dotsb\otimes v'_n$ and $v_i\in V_i$, $v'_i\in V_i$ are such that there are at least two linearly independent pairs of vectors among the pairs $\{(v_i,v'_i)\}$ for each $i$.
(The EPR state for $n=2$ and the GHZ state~\cite{GHZ} for $n=3$ are such examples.)
It seems plausible (and will be proved later) that linear combinations with a larger number of terms and a smaller number of linear relations among vectors in each tensor product represent states with larger degrees of entanglement.
To define degrees of entanglement, we proceed as follows.

We first note that a degree of entanglement is an algebraic invariant: a quantity that depends only on properties of spaces and does not depend on properties of individual vectors.
Here such invariants can appear only as dimensions of linear subspaces of $V$, and only subspaces linearly depending on $v$ can participate in classification of entangled states.
To find the required subspaces, we note that any linear subspace can be defined using an appropriate linear map.
Specifically, vector spaces $W$ and $W'$ together with a linear map $f:W\to W'$ define two associated fundamental subspaces: the kernel and image of the map,
\begin{align*}
  \ker{f} &= \{w\in W\colon f(w)=0\}\subset W, \\
  \im{f} &= \{w'\in W'\colon w'=f(w), w\in W\}\subset W'.
\end{align*}
Introducing the transpose map $f'\colon W'\to W$, we find that the matrices of $f$ and $f'$ are the transposes of each other, which implies that the kernels and images of $f$ and $f'$ are related.
Thus, if both maps are used to classify subspaces, then it suffices to consider only their kernels, for example.

Since we seek a map $f(v)\colon W\to W'$ that is linear in $v$, we have to choose $f(v)(w)=v\otimes w^*$, where $W$, $W'$ are such that $W\otimes W'=V$ and $w^*$ is the dual of $w$.
The kernel $K(v)=\ker{f(v)}$ and the invariant $k(v)=\dim{K(v)}$ describe an entanglement property of $v$ associated with a particular choice of $(W,W')$.
For $w\in K(v)$, the equation $v\otimes w^*=0$ implies the general form of $v$,
\begin{align*}
  v=\sum_{i=1}^{\dim{W}-k(v)} w_i\otimes w'_i, \quad \{w_i\}\subset W, \quad \{w'_i\}\subset W', \\
  \dim{\linearspan{(\{w_i\})}}=\dim{\linearspan{(\{w'_i\})}}=\dim{W}-k(v),
\end{align*}
which is a convenient computational tool that allows us to group states with identical entanglement properties.

To complete the formulation of the method, we need to explore all combinatorial possibilities present in the problem.
To this end, we choose all possible pairs of spaces $(W,W')$ such that $W\otimes W'=V$, and for each such choice we find the corresponding map $f(v)$, the kernel $K(v)$, and the invariant $k(v)$ for each $v\in V$.
The resulting set of the kernels $\{K(v)\}$ determines the entanglement class of $v$.
It turns out, however, that the sequence (ordered set) of the invariants $(k(v))$ does not suffice to specify the class uniquely, but the sequence of the dimensions of all possible intersections of elements of $\{\ker{(f(v)\otimes\id_{W'})}\}$ does.
Now choosing the smallest subsequence of independent invariants among such a sequence, we arrive at the complete set of degrees of entanglement of elements of $V$.
Finally, by examining possible values of the invariants, we find the set of entanglement classes of $V$.

\section{Examples}

Our classification method works for arbitrary finite $n$ and $D=(\dim{V_i})_{1\le i\le n}$.
As illustrative examples, we consider entanglement for the case $n=2$ for arbitrary $D=(d_1,d_2)$ and the case $n=3$ for $D=(2,2,d)$ and $D=(2,3,d)$, where $d$ is arbitrary and corresponds to spin $\frac{1}{2}(d-1)$.

We first consider the case $n=2$, $D=(d_1,d_2)$.
Maps and associated quantities are given in Table~\ref{table:maps2}.
The invariants $k_1(v)$, $k_2(v)$ describe properties of $v$ that are related to the partition of the system $S$ into subsystems $(S_1,S_2)$, $(S_2,S_1)$, respectively.
Since the partitions $(S_1,S_2)$, $(S_2,S_1)$ are equivalent, there is a relation between $k_1(v)$, $k_2(v)$; specifically $d_1-k_1(v)=d_2-k_2(v)$ for each $v\in V$.
This results in the complete set of entanglement classes and general forms of their elements given in Table~\ref{table:d1d2}.

\begin{table}[ht]
  \caption{\label{table:maps2} Maps and associated quantities for $n=2$, $D=(d_1,d_2)$.}
  \begin{ruledtabular}
    \begin{tabular}{llll}
      $(W,W')$    & $f(v)$   & $K(v)$   & $k(v)$ \\ \hline
      $(V_1,V_2)$ & $f_1(v)$ & $K_1(v)$ & $k_1(v)$ \\
      $(V_2,V_1)$ & $f_2(v)$ & $K_2(v)$ & $k_2(v)$
    \end{tabular}
  \end{ruledtabular}
\end{table}

\begin{table}[ht]
  \caption{\label{table:d1d2} The entanglement classes $\{C_l\}_{0\le l\le m}$, $m=\min{\{d_1,d_2\}}$, their independent algebraic invariants $k_1(v)$, and general forms of their representative elements for $n=2$, $D=(d_1,d_2)$.
  Each expression $[j,j]$ stands for $u_{1,j}\otimes u_{2,j}$, where $\{u_{i,j}\}$ is a set of any linearly independent elements of $V_i$.
  (For example, a representative element of $C_2$ is the EPR state $v=u_{1,1}\otimes u_{2,1}+u_{1,2}\otimes u_{2,2}$.)}
  \begin{ruledtabular}
    \begin{tabular}{lll}
      & $k_1(v)$ & $v$ \\ \hline
      $C_0$ & $d_1$ & $0$ \\
      $C_l$ & $d_1-l$ & $[1,1]+\cdots+[l,l]$ \\
    \end{tabular}
  \end{ruledtabular}
\end{table}

Next we consider the case $n=3$, $D=(d_1,d_2,d_3)$.
Maps and associated quantities are given in Table~\ref{table:maps3}.
The invariants $k_1(v)$, $k_2(v)$, $k_3(v)$, $k_{1,2}(v)$, $k_{1,3}(v)$, $k_{2,3}(v)$ describe properties of $v$ that are related to the partition of the system $S$ into subsystems $(S_1,S_2\cup S_3)$, $(S_2,S_1\cup S_3)$, $(S_3,S_1\cup S_2)$, $(S_1\cup S_2,S_3)$, $(S_1\cup S_3,S_2)$, $(S_2\cup S_3,S_1)$, respectively.
Similarly to the case of two spaces, the partitions in each group in $((S_1,S_2\cup S_3),(S_2\cup S_3,S_1))$, $((S_2,S_1\cup S_3),(S_1\cup S_3,S_2))$, $((S_3,S_1\cup S_2),(S_1\cup S_2,S_3))$ are equivalent and there are relations 
\begin{align*}
  d_1-k_1(v)=d_2 d_3-k_{2,3}(v), \ d_2-k_2(v)=d_1 d_3-k_{1,3}(v), \ d_3-k_3(v)=d_1 d_2-k_{1,2}(v)
\end{align*}
for each $v\in V$.

\begin{table}[ht]
  \caption{\label{table:maps3} Maps and associated quantities for $n=3$, $D=(d_1,d_2,d_3)$.}
  \begin{ruledtabular}
    \begin{tabular}{llll}
      $(W,W')$               & $f(v)$       & $K(v)$       & $k(v)$ \\ \hline
      $(V_1,V_2\otimes V_3)$ & $f_1(v)$     & $K_1(v)$     & $k_1(v)$ \\
      $(V_2,V_1\otimes V_3)$ & $f_2(v)$     & $K_2(v)$     & $k_2(v)$ \\
      $(V_3,V_1\otimes V_2)$ & $f_3(v)$     & $K_3(v)$     & $k_3(v)$ \\
      $(V_1\otimes V_2,V_3)$ & $f_{1,2}(v)$ & $K_{1,2}(v)$ & $k_{1,2}(v)$ \\
      $(V_1\otimes V_3,V_2)$ & $f_{1,3}(v)$ & $K_{1,3}(v)$ & $k_{1,3}(v)$ \\
      $(V_2\otimes V_3,V_1)$ & $f_{2,3}(v)$ & $K_{2,3}(v)$ & $k_{2,3}(v)$
    \end{tabular}
  \end{ruledtabular}
\end{table}

When classifying states for $n=3$, we find a feature not present for $n=2$.
Namely, states with the same values of the invariants $k_1(v)$, $k_2(v)$, $k_3(v)$ can be distinguished with the help of an additional subspace of $V$,
\begin{align*}
  K_{1,2,3}(v) &= \ker{(f_{1,2}\otimes\id_3)}\cap\ker{(f_{1,3}\otimes\id_2)}\cap\ker{(f_{2,3}\otimes\id_1)}, 
  \label{K123}
\end{align*} 
where $\id_i\colon V_i\to V_i$ is the identity map. 
The invariant $k_{1,2,3}(v)=\dim{K_{1,2,3}(v)}$ describes a property of $v$ that is related to the partition of the system $S$ into subsystems $(S_1,S_2,S_3)$, and $k_{1,2,3}(v)$ is irreducible in the sense that it cannot be expressed in terms of the invariants $k_1(v)$, $k_2(v)$, $k_3(v)$.

\begin{table}[ht]
  \caption{\label{table:22d} The entanglement classes, their algebraic invariants, and general forms of their representative elements for $n=3$, $D=(2,2,d)$.
  Classes for which any of the invariants in $(k_3(v),k_{1,2,3}(v))$ are negative should be discarded.
  Classes within a horizontal block are added each time $d$ increases by $1$, so that there are $7$, $9$, $10$ classes for $d=2$, $d=3$, $d\ge 4$, respectively.
  Each expression $[j_1,j_2,j_3]$ stands for $u_{1,j_1}\otimes u_{2,j_2}\otimes u_{3,j_3}$, where $\{u_{i,j}\}$ is a set of any linearly independent elements of $V_i$.
  (For example, for $d=2$, the GHZ state is $v=[1,1,1]+[2,2,2]=u_{1,1}\otimes u_{2,1}\otimes u_{3,1}+u_{1,2}\otimes u_{2,2}\otimes u_{3,2}$ in this notation.)}
  \begin{ruledtabular}
    \begin{tabular}{llllll}
      & $k_1(v)$ & $k_2(v)$ & $k_3(v)$ & $k_{1,2,3}(v)$ & $v$ \\ \hline
      $C_0$ & $2$ & $2$ & $d$   & $4d$   & $0$ \\ \hline
      $C_1$ & $1$ & $1$ & $d-1$ & $3d-2$ & $[1,1,1]$ \\
      $C_2$ & $0$ & $0$ & $d-1$ & $3d-3$ & $[1,1,1]+[2,2,1]$ \\ \hline
      $C_3$ & $0$ & $1$ & $d-2$ & $2d-1$ & $[1,1,1]+[2,1,2]$ \\
      $C_4$ & $1$ & $0$ & $d-2$ & $2d-1$ & $[1,1,1]+[1,2,2]$ \\
      $C_5$ & $0$ & $0$ & $d-2$ & $2d-3$ & $[1,1,1]+[1,2,2]+[2,1,2]$ \\
      $C_6$ & $0$ & $0$ & $d-2$ & $2d-4$ & $[1,1,1]+[2,2,2]$ \\ \hline
      $C_7$ & $0$ & $0$ & $d-3$ & $d-2$  & $[1,1,1]+[1,2,2]+[2,2,3]$ \\
      $C_8$ & $0$ & $0$ & $d-3$ & $d-3$  & $[1,1,1]+[1,2,2]+[2,1,2]+[2,2,3]$ \\ \hline
      $C_9$ & $0$ & $0$ & $d-4$ & $0$    & $[1,1,1]+[1,2,2]+[2,1,3]+[2,2,4]$ \\
    \end{tabular}
  \end{ruledtabular}
\end{table}
\begin{table}[ht]
  \caption{\label{table:23d} The entanglement classes, their algebraic invariants, and general forms of their representative elements for $n=3$, $D=(2,3,d)$.
  Classes for which any of the invariants in $(k_3(v),k_{1,2,3}(v))$ are negative should be discarded.
  Classes within a horizontal block are added each time $d$ increases by $1$, so that there are $9$, $17$, $23$, $25$, $26$ classes for $d=2$, $d=3$, $d=4$, $d=5$, $d\ge 6$, respectively.
  Each expression $[j_1,j_2,j_3]$ stands for $u_{1,j_1}\otimes u_{2,j_2}\otimes u_{3,j_3}$, where $\{u_{i,j}\}$ is a set of any linearly independent elements of $V_i$.}
  \begin{ruledtabular}
    \begin{tabular}{llllll}
      & $k_1(v)$ & $k_2(v)$ & $k_3(v)$ & $k_{1,2,3}(v)$ & $v$ \\ \hline
      $C_0$    & $2$ & $3$ & $d$   & $6d$   & $0$ \\ \hline
      $C_1$    & $1$ & $2$ & $d-1$ & $5d-3$ & $[1,1,1]$ \\
      $C_2$    & $0$ & $1$ & $d-1$ & $5d-5$ & $[1,1,1]+[2,2,1]$ \\ \hline
      $C_3$    & $0$ & $2$ & $d-2$ & $4d-2$ & $[1,1,1]+[2,1,2]$ \\
      $C_4$    & $1$ & $1$ & $d-2$ & $4d-3$ & $[1,1,1]+[1,2,2]$ \\
      $C_5$    & $0$ & $1$ & $d-2$ & $4d-5$ & $[1,1,1]+[1,2,2]+[2,1,2]$ \\
      $C_6$    & $0$ & $1$ & $d-2$ & $4d-6$ & $[1,1,1]+[2,2,2]$ \\
      $C_7$    & $0$ & $0$ & $d-2$ & $4d-7$ & $[1,1,1]+[1,2,2]+[2,3,1]$ \\
      $C_8$    & $0$ & $0$ & $d-2$ & $4d-8$ & $[1,1,1]+[1,2,2]+[2,2,1]+[2,3,2]$ \\ \hline
      $C_9$    & $1$ & $0$ & $d-3$ & $3d-1$ & $[1,1,1]+[1,2,2]+[1,3,3]$ \\
      $C_{10}$ & $0$ & $1$ & $d-3$ & $3d-4$ & $[1,1,1]+[1,2,2]+[2,1,3]$ \\
      $C_{11}$ & $0$ & $1$ & $d-3$ & $3d-5$ & $[1,1,1]+[1,2,2]+[2,1,2]+[2,2,3]$ \\
      $C_{12}$ & $0$ & $0$ & $d-3$ & $3d-5$ & $[1,1,1]+[1,2,2]+[1,3,3]+[2,1,2]$ \\
      $C_{13}$ & $0$ & $0$ & $d-3$ & $3d-6$ & $[1,1,1]+[1,2,2]+[2,3,3]$ \\
      $C_{14}$ & $0$ & $0$ & $d-3$ & $3d-7$ & $[1,1,1]+[1,2,2]+[1,3,3]+[2,1,2]+[2,2,3]$ \\
      $C_{15}$ & $0$ & $0$ & $d-3$ & $3d-8$ & $[1,1,1]+[1,2,2]+[2,1,3]+[2,3,1]$ \\
      $C_{16}$ & $0$ & $0$ & $d-3$ & $3d-9$ & $[1,1,1]+[1,2,2]+[2,2,2]+[2,3,3]$ \\ \hline
      $C_{17}$ & $0$ & $1$ & $d-4$ & $2d-2$ & $[1,1,1]+[1,2,2]+[2,1,3]+[2,2,4]$ \\
      $C_{18}$ & $0$ & $0$ & $d-4$ & $2d-3$ & $[1,1,1]+[1,2,2]+[1,3,3]+[2,3,4]$ \\
      $C_{19}$ & $0$ & $0$ & $d-4$ & $2d-5$ & $[1,1,1]+[1,2,2]+[1,3,3]+[2,2,4]+[2,3,1]$ \\
      $C_{20}$ & $0$ & $0$ & $d-4$ & $2d-6$ & $[1,1,1]+[1,2,2]+[2,2,3]+[2,3,4]$ \\
      $C_{21}$ & $0$ & $0$ & $d-4$ & $2d-7$ & $[1,1,1]+[1,2,2]+[1,3,3]+[2,2,3]+[2,3,4]$ \\
      $C_{22}$ & $0$ & $0$ & $d-4$ & $2d-8$ & $[1,1,1]+[1,2,2]+[1,3,3]+[2,1,2]+[2,2,3]+[2,3,4]$ \\ \hline
      $C_{23}$ & $0$ & $0$ & $d-5$ & $d-3$  & $[1,1,1]+[1,2,2]+[1,3,3]+[2,1,4]+[2,2,5]$ \\
      $C_{24}$ & $0$ & $0$ & $d-5$ & $d-5$  & $[1,1,1]+[1,2,2]+[1,3,3]+[2,1,3]+[2,2,4]+[2,3,5]$ \\ \hline
      $C_{25}$ & $0$ & $0$ & $d-6$ & $0$    & $[1,1,1]+[1,2,2]+[1,3,3]+[2,1,4]+[2,2,5]+[2,3,6]$
    \end{tabular}
  \end{ruledtabular}
\end{table}

As our main computational device, we use the general forms of $v$ that follow from the equation $v\otimes w^*=0$ for three cases $w\in K_1(v)$, $w\in K_2(v)$, $w\in K_3(v)$.
The consistency of the resulting forms leads to restrictions on possible values of invariants and consequently to the complete set of equivalent classes.

Although it is possible to perform these computations for any $D$, we give the results only for the cases $D=(2,2,d)$ and $D=(2,3,d)$, where $d$ is arbitrary; other cases are similarly treated.
Due to the special role played by $V_3$ in these examples, it is convenient to proceed by first considering each possible value of $k_3(v)$, then finding allowed values of $k_1(v)$, $k_2(v)$, and finally those of $k_{1,2,3}(v)$.
As a result, we obtain the complete set of entanglement classes and general forms of their reperesentative elements as given in Tables~\ref{table:22d} and \ref{table:23d}.
For each class in these tables, there are several possible general forms of reperesentative elements related by certain permutations; see Appendix for details of the case $D=(2,2,2)$ ($3$ qubits) and \cite{Buniy:2010zp} for further details.

\section{Conclusions}

The superposition principle and the tensor product postulate in quantum mechanics give rise to the phenomenon of entanglement.
As a result, a state vector of a system consisting of several subsystems is a linear combination of tensor products of state vectors of the subsystems.
The nature of the linear combination determines the entanglement of the state vector.
More specifically, properties of algebraic structures associated with states can be used to derive entanglement invariants.

We developed a method of classification of entangled states that uses linear maps to define degrees of entanglement.
Our classification uses discrete algebraic invariants, which should be contrasted with the standard continuous invariants.

For cases found in the literature, entanglement classifications obtained by using our method coincide with results obtained by other methods.
We also obtained results for cases that were either unsolved or only conjectured in the literature.
In particular, for the case $n=3$, $D=(2,2,d)$ for $d=2,\dotsc,5$, our method gives the same number of classes as classifications in \cite{Gelfand}, \cite{Dur}, \cite{Miyake1}, \cite{Miyake2}, while for $d>5$, our method gives the same number of classes as the conjectured classification in \cite{Miyake1}, \cite{Miyake2}.
Our entanglement classes and representative elements for $D=(2,3,d)$ are all new.

Although we considered here only some of the simpler cases of three subsystems, other cases are only slightly more complicated.
In a further study~\cite{Buniy:2010zp}, we consider a large selection of such cases and formulate a general conjecture about entanglement classes for all cases of three subsystems.
Using our method, we also obtain entanglement classes of four qubits~\cite{Buniy:2010zp}.
Furthermore, we believe the complete classification of entanglement of five qubits is now within reach and we plan to study it in the near future.

\begin{acknowledgments}
We thank Mike Duff for a useful discussion and encouragement.
RVB acknowledges support from DOE grant at ASU and from Arizona State Foundation.
The work of TWK was supported by DOE grant number DE-FG05-85ER40226 and he thanks the Aspen Center for Physics for hospitality while this work was in progress.
\end{acknowledgments}

\appendix*
\section{Three qubits}
For an arbitrary vector $v\in V_1\otimes V_2\otimes V_3$ in its general form
\begin{align*}
  v&=\sum_{i=1}^2\sum_{j=1}^2\sum_{k=1}^2 v_{ijk}e_{1,i}\otimes e_{2,j}\otimes e_{3,k}, 
\end{align*}
the linear maps $f(v)$ are given by
\begin{align*}
  f_1(v)(w)&=\sum_{i=1}^2\sum_{j=1}^2\sum_{k=1}^2 v_{ijk}w_i e_{2,j}\otimes e_{3,k},\ w\in V_1,\\ 
  f_2(v)(w)&=\sum_{i=1}^2\sum_{j=1}^2\sum_{k=1}^2 v_{ijk}w_j e_{1,i}\otimes e_{3,k},\ w\in V_2,\\ 
  f_3(v)(w)&=\sum_{i=1}^2\sum_{j=1}^2\sum_{k=1}^2 v_{ijk}w_k e_{1,i}\otimes e_{2,j},\ w\in V_3,\\
  f_{1,2}(v)(w)&=\sum_{i=1}^2\sum_{j=1}^2\sum_{k=1}^2 v_{ijk}w_{ij}e_{3,k},\ w\in V_1\otimes V_2,\\ 
  f_{1,3}(v)(w)&=\sum_{i=1}^2\sum_{j=1}^2\sum_{k=1}^2 v_{ijk}w_{ik}e_{2,j},\ w\in V_1\otimes V_3,\\ 
  f_{2,3}(v)(w)&=\sum_{i=1}^2\sum_{j=1}^2\sum_{k=1}^2 v_{ijk}w_{jk}e_{1,i},\ w\in V_2\otimes V_3.
\end{align*}
The associated kernels $K(v)=\ker{f(v)}$ are
\begin{align*}
  K_1(v) &= \{w\in V_1 \colon \sum_{i=1}^2 v_{ijk}w_i=0, \ j,k\in\{1,2\}\}, \\
  K_2(v) &= \{w\in V_2 \colon \sum_{j=1}^2 v_{ijk}w_j=0, \ i,k\in\{1,2\}\}, \\
  K_3(v) &= \{w\in V_3 \colon \sum_{k=1}^2 v_{ijk}w_k=0, \ i,j\in\{1,2\}\}, \\
  K_{1,2}(v) &= \{w\in V_1\otimes V_2 \colon \sum_{i=1}^2\sum_{j=1}^2 v_{ijk}w_{ij}=0, \ k\in\{1,2\}\}, \\
  K_{1,3}(v) &= \{w\in V_1\otimes V_3 \colon \sum_{i=1}^2\sum_{k=1}^2 v_{ijk}w_{ik}=0, \ j\in\{1,2\}\}, \\
  K_{2,3}(v) &= \{w\in V_2\otimes V_3 \colon \sum_{j=1}^2\sum_{k=1}^2 v_{ijk}w_{jk}=0, \ i\in\{1,2\}\}.
\end{align*}
The constraint equations for the above kernels follows directly from their definitions.
The slightly more complicated kernel $K_{1,2,3}(v)$ (with the resulting $12$ constraint equations) follows from its definition,
\begin{align*}
  K_{1,2,3}(v) = \Bigl\{w\in V_1\otimes V_2\otimes V_3 \colon \sum_{i=1}^2\sum_{j=1}^2 v_{ijk}w_{ijl}=0, \ k,l\in\{1,2\}; \\ \sum_{i=1}^2\sum_{k=1}^2 v_{ijk}w_{ilk}=0, \ j,l\in\{1,2\}; \ \ \sum_{j=1}^2\sum_{k=1}^2 v_{ijk}w_{ljk}=0, \ i,l\in\{1,2\}\Bigr\},
\end{align*}
which  we leave for the reader as an exercise.

Our task now is to find all allowed values for the dimensions of the kernels among their possible values given by
\begin{align*}
  0\le\dim{K_i(v)}\le 2, & \ i\in\{1,2,3\},\\
  0\le\dim{K_{j,k}(v)}\le 4, & \ (j,k)\in\{(1,2),(1,3),(2,3)\},\\
  0\le\dim{K_{1,2,3}(v)}\le 8.
\end{align*}
We consider the following cases:

Case 1: Let $\dim{K_1(v)}=2$.
This means that $w$ in the definition of $K_1(v)$ is arbitrary.
Since $w$ satisfies $4$ constraint equations, we find that the equations are over-constrained unless $v=0$. 
This leads to
\begin{align*}
  \dim{K_1(v)}=\dim{K_2(v)}=\dim{K_3(v)}&=2,\\
  \dim{K_{1,2}(v)}=\dim{K_{1,3}(v)}=\dim{K_{2,3}(v)}&=4,\\
  \dim{K_{1,2,3}(v)}&=8
\end{align*}
and gives the class $C_0$.
  
Case 2: Let $\dim{K_1(v)}=1$, which implies $v\not=0$. 
This means that $w$ in the definition of $K_1(v)$ satisfies 
\begin{align*}
  c_1w_1+c_2w_2=0
\end{align*}
for some constants $c_1$ and $c_2$.
Solving for $w_2$ and substituting into the $K_1(v)$ equations leads to 
\begin{align*}
  \frac{v_{211}}{v_{111}}=\frac{v_{212}}{v_{112}}=\frac{v_{221}}{v_{121}}=\frac{v_{222}}{v_{122}}.
\end{align*}
Using these results in the $K_2(v)$ equations, we find
\begin{align*}
  v_{111}w_1+v_{121}w_2=0,\\
  v_{112}w_1+v_{122}w_2=0.
\end{align*}
This allows three possibilities:

Case 2.1: Let $\dim{K_2(v)}=2$.
This implies $v_{111}=v_{121}=v_{112}=v_{122}=0$, which contradicts the requirement $v\not=0$.

Case 2.2: Let $\dim{K_2(v)}=1$.
This implies
\begin{align*}
  c'_1w_1 +c'_2w_2=0
\end{align*}
for some constants $c'_1$ and $c'_2$.
Using this equation in the remaining two $K_2(v)$ equations, we find
\begin{align*}
  \frac{v_{121}}{v_{111}}=\frac{v_{122}}{v_{112}}.
\end{align*}
Using these results in the $K_3(v)$ equations, we find they reduce to a single independent equation
\begin{align*}
  v_{111}w_1+v_{112}w_2=0.
\end{align*}
Hence $\dim{K_3(v)}=1$.
Now using the above results in the $K_{1,2}(v),K_{1,3}(v),K_{2,3}(v)$, we find that there is only one constraint for each of these kernels,
\begin{align*}
  v_{111}^2w_{11}+v_{111}v_{121}w_{12}+v_{111}v_{211}w_{21}+v_{121}v_{211}w_{22}&=0, \ w\in K_{1,2}(v),\\
  v_{111}^2w_{11}+v_{111}v_{112}w_{12}+v_{111}v_{211}w_{21}+v_{112}v_{211}w_{22}&=0, \ w\in K_{1,3}(v),\\
  v_{111}^2w_{11}+v_{111}v_{112}w_{12}+v_{111}v_{121}w_{21}+v_{112}v_{121}w_{22}&=0, \ w\in K_{2,3}(v).
\end{align*}
Since there is one constraint for each $4$-dimensional vector $w$, we conclude $\dim{K_{1,2}(v)}=\dim{K_{1,3}(v)}=\dim{K_{2,3}(v)}=3$. 

Finally we turn to $K_{1,2,3}(v)$. After some straightforward algebra, $12$ equations in the definition of $K_{1,2,3}(v)$ reduce to 
\begin{align*}
  &w_{112}=-\frac{1}{v_{111}^2}(v_{111}v_{121}w_{122}+v_{111}v_{211}w_{212}+v_{121}v_{211}w_{222}),\\
  &w_{121}=-\frac{1}{v_{111}^2}(v_{111}v_{112}w_{122}+v_{111}v_{211}w_{221}+v_{112}v_{211}w_{222}),\\
  &w_{211}=-\frac{1}{v_{111}^2}(v_{111}v_{112}w_{212}+v_{111}v_{121}w_{221}+v_{112}v_{121}w_{222}),\\
  &v_{111}^2w_{111}-v_{112}v_{121}w_{122}-v_{112}v_{211}w_{212}-v_{121}v_{211}w_{221}-2\frac{v_{112}v_{121}v_{211}}{v_{111}}w_{222}=0,
\end{align*}
from which we read off $\dim{K_{1,2,3}(v)}=4$.
This is the class $C_1$.

Case 2.3: Let $\dim{K_2(v)}=0$.
Algebra similar to what we have encountered above leads to the result $\dim{K_3(v)}=0$, $\dim{K_{1,2}(v)}=2$, $\dim{K_{1,3}(v)}=2$, $\dim{K_{2,3}(v)}=3$, and
$\dim{K_{1,2,3}(v)}=3$.
There are also two more Cases $2.3'$ and $2.3''$, where we cyclically permute indices $1$, $2$ and $3$.
We obtain the classes $C_2,C_3,C_4$.

Case 3: Here a similar analysis shows that we can have $\dim{K_1(v)}=\dim{K_2(v)}=\dim{K_3(v)}=0$ and $\dim{K_{1,2}(v)}=\dim{K_{1,3}(v)}=\dim{K_{2,3}(v)}=2$ with two subcases.

Case $3.1$: $\dim{K_{1,2,3}(v)}=1$, which is the class $C_5$.

Case $3.2$: $\dim{K_{1,2,3}(v)}=0$, which is the class $C_6$.

This completes the list of the full set of allowed values of the linear invariants and the resulting equivalence classes for $3$ qubits.
These results are summarized in Table \ref{table:22d} when we set $d=2$ there.

\end{document}